\documentclass[lettersize,journal]{IEEEtran}
\usepackage{amsmath,amsfonts}
\usepackage{array}
\usepackage{textcomp}
\usepackage{stfloats}
\usepackage{url}
\usepackage{verbatim}
\usepackage{graphicx}
\usepackage{cite}
\usepackage{cite}
\usepackage{microtype}
\usepackage{hyperref}
\usepackage{url}
\usepackage{booktabs}
\usepackage{xspace}
\usepackage{caption}
\usepackage{subcaption}

\usepackage{float}
\usepackage{enumitem}
\usepackage{pifont}
\usepackage{latexsym}

\usepackage{xcolor}
\usepackage{colortbl}

\definecolor{lightblue}{RGB}{224, 235, 255}
\definecolor{mediumblue}{RGB}{180, 205, 255}

\usepackage{graphicx}

\usepackage{listings}

\usepackage{microtype}

\usepackage{inconsolata}

\usepackage{graphicx}

\usepackage{multirow}
\usepackage{hhline}

\usepackage{lineno}

\usepackage{amsmath,amssymb,amsfonts}
\usepackage{cleveref}
\usepackage[most]{tcolorbox}
\usepackage{graphicx}
\usepackage{textcomp}
\usepackage{xcolor}

\usepackage{algorithm}
\usepackage{algorithmicx}
\usepackage[noend]{algpseudocode}

\usepackage{tabularx}
\usepackage{orcidlink}

\begin{document}

\newcommand{\name}{VeriCoder\xspace}

\newcommand{\numdataset}{xxx\xspace}

\newcommand{\cmark}{\textcolor{green!60!black}{\ding{51}}}
\newcommand{\xmark}{\textcolor{red!80!black}{\ding{55}}}

\newcommand{\basemodel}{Qwen2.5-14B-Instruct\xspace}

\newcommand{\veriloghumanpassone}{38.3\%\xspace}
\newcommand{\verilogmachinepassone}{55.7\%\xspace}

\newcommand{\rtllmpass}{48.3\%\xspace}

\newcommand{\anjiang}[1]{\textcolor{green}{\textbf{Anjiang:} #1}}
\newcommand{\huanmi}[1]{\textcolor{blue}{\textbf{Huanmi:} #1}}

\newcommand{\ke}[1]{\textcolor{red}{\textbf{Ke:} #1}}

\newcommand{\caroline}[1]{\textcolor{magenta}{\textbf{Caroline:} #1}}

\newcommand{\yao}[1]{\textcolor{purple}{\textbf{yao:} #1}}

\newcommand{\daniel}[1]{\textcolor{purple}{\textbf{Daniel:} #1}}

\newcommand{\thiago}[1]{\textcolor{yellow}{\textbf{thiago:} #1}}

\newcommand{\tarun}[1]{\textcolor{orange}{\textbf{Tarun:} #1}}

\newcommand{\alex}[1]{\textcolor{red}{\textbf{Alex:} #1}}

\newcommand{\CodeIn}[1]{\texttt{#1}}

\definecolor{veriloggray}{rgb}{0.95,0.95,0.95}
\definecolor{verilogblue}{rgb}{0.2,0.2,0.6}
\definecolor{verilogred}{rgb}{0.6,0.0,0.0}
\definecolor{veriloggreen}{rgb}{0.0,0.5,0.0}

\lstdefinelanguage{Verilog}{
  morekeywords={
    module, endmodule, input, output, reg, wire, always, begin, end,
    if, else, for, assign, case, endcase, default, parameter, localparam,
    integer
  },
  sensitive=true,
  morecomment=[l]{//},
  morecomment=[s]{/*}{*/},
  morestring=[b]",
}

\lstdefinestyle{verilog-style}{
  language=Verilog,
  backgroundcolor=\color{veriloggray},
  basicstyle=\ttfamily\footnotesize,
  keywordstyle=\color{verilogblue}\bfseries,
  commentstyle=\color{veriloggreen}\itshape,
  stringstyle=\color{verilogred},
  numbers=left,
  numberstyle=\tiny\color{gray},
  numbersep=4pt,              
  xleftmargin=2.5em,          
  framexleftmargin=1em,       
  xrightmargin=1em,
  tabsize=2,
  showstringspaces=false,
  breaklines=true,
  captionpos=b,
  frame=single,
  rulecolor=\color{black},
  literate={`}{{\textasciigrave}}1,
}

\definecolor{promptframe}{rgb}{0.2, 0.6, 0.6}  
\definecolor{promptbg}{rgb}{0.97, 0.98, 0.98}  
\definecolor{prompttitlebg}{rgb}{0.88, 0.94, 0.94}  

\definecolor{templateframe}{rgb}{0.45, 0.3, 0.6}     
\definecolor{templatebg}{rgb}{0.97, 0.96, 0.99}       
\definecolor{templatetitlebg}{rgb}{0.88, 0.85, 0.95}  

\newtcolorbox{templatebox}{
  colback       = templatebg,
  colframe      = templateframe,
  coltitle      = black,
  colbacktitle  = templatetitlebg,
  fonttitle     = \bfseries,
  title         = Prompt Template,
  fontupper     = \footnotesize,
  boxrule       = 0.6pt,
  arc           = 1.5mm,
  left          = 2mm,
  right         = 2mm,
  top           = 1mm,
  bottom        = 1mm,
}

\newtcolorbox{promptbox}{
  colback       = promptbg,
  colframe      = promptframe,
  coltitle      = black,
  colbacktitle  = prompttitlebg,
  fonttitle     = \bfseries,
  title         = Natural Language Specification,
  fontupper   = \footnotesize, 
  boxrule       = 0.6pt,
  arc           = 1.5mm,
  left          = 2mm,
  right         = 2mm,
  top           = 1mm,
  bottom        = 1mm,
}

\title{VeriCoder: Enhancing LLM-Based RTL Code Generation through Functional Correctness Validation}


\author{

Anjiang Wei\,\orcidlink{0000-0003-1654-6027}, %
Huanmi Tan\,\orcidlink{0009-0006-2710-1532},%
Tarun Suresh\,\orcidlink{0000-0002-1426-7633}, %
Daniel Mendoza\,\orcidlink{0009-0006-3431-6566}, \\
Thiago S. F. X. Teixeira\,\orcidlink{0000-0002-8031-0652}, %
Ke Wang\,\orcidlink{0000-0003-0844-5023},%
Caroline Trippel\,\orcidlink{0000-0002-5776-1121}, %
and Alex Aiken\,\orcidlink{0000-0002-3723-9555}%

\thanks{
Anjiang Wei, Daniel Mendoza, Caroline Trippel, and Alex Aiken are affiliated with Stanford University (e-mail: anjiang@cs.stanford.edu; dmendo@stanford.edu; trippel@stanford.edu; aiken@cs.stanford.edu).
}
\thanks{
Huanmi Tan is affiliated with Carnegie Mellon University (e-mail: huanmi.tan@gmail.com).
}
\thanks{
Tarun Suresh is affiliated with University of Illinois Urbana-Champaign (e-mail: tsuresh3@illinois.edu).
}
\thanks{
Thiago S. F. X. Teixeira is with Intel Corporation (e-mail: thiago.teixeira@intel.com).
}
\thanks{
Ke Wang is with Nanjing University (e-mail: kwg@nju.edu.cn).
}
}


\markboth{}%
{Wei \MakeLowercase{\textit{et al.}}: VeriCoder: Enhancing LLM-Based RTL Code Generation through Functional Correctness Validation}


\maketitle

\begin{abstract}
Recent advances in Large Language Models (LLMs) have sparked growing interest in applying them to Electronic Design Automation (EDA) tasks, particularly Register Transfer Level (RTL) code generation. While several RTL datasets have been introduced, most focus on syntactic validity rather than functional validation with tests, leading to training examples that compile but may not implement the intended behavior. We present \textsc{VeriCoder}, a model for RTL code generation fine-tuned on a dataset validated for functional correctness. This fine-tuning dataset is constructed using a novel methodology that combines unit test generation with feedback-directed refinement. Given a natural language specification and an initial RTL design, we prompt a teacher model (GPT-4o-mini) to generate unit tests and iteratively revise the 
RTL design based on its simulation results using the generated tests. If necessary, the teacher model also updates the tests to ensure they comply with the natural language specification. As a result of this process, every example in our dataset is functionally validated, consisting of a natural language description, an RTL implementation, and passing tests. Fine-tuned on this dataset of 125,777 examples, \textsc{VeriCoder} achieves state‑of‑the‑art metrics in functional correctness on VerilogEval and RTLLM, with relative gains of up to 71.7\% and 27.4\%, respectively. An ablation study further shows that models trained on our functionally validated dataset outperform those trained on functionally non-validated datasets, underscoring the importance of high-quality datasets in RTL code generation. Our code, data, and models are publicly available at \url{https://github.com/Anjiang-Wei/VeriCoder}
\end{abstract}


\section{Introduction}
\label{sec:intro}

Large Language Models (LLMs) have demonstrated remarkable performance across natural language processing tasks, spurring growing interest in applying their capabilities to a broad range of Electronic Design Automation (EDA) problems~\cite{liu2023chipnemo,chen2024dawn,zhong2023llm4eda,he2024large}. Recent efforts explore LLMs for code generation~\cite{yao2024rtlrewriter,liu2024rtlcoder,liu2024rtlcoder2,cui2024origen,liu2023verilogeval,lu2024rtllm,tsai2024rtlfixer,liao2024llms}, architecture design~\cite{fu2023gpt4aigchip,yan2023viability,liang2023unleashing}, verification~\cite{cosler2023nl2spec,sun2023towards}, tool assistance~\cite{wu2024chateda,xiao2025eda}, and debugging~\cite{liu2023chipnemo,xu2024meic}. In this work, we focus on generating Register Transfer Level (RTL) code from natural language specifications. Automating RTL code generation has the potential to significantly boost hardware design productivity and reduce the manual effort involved in complex design tasks, making it a timely and impactful area of research.

Developing open-source, lightweight models for RTL code generation is essential for advancing both research and deployment. Proprietary models such as GPT-4o and Claude 3.7 restrict customization and lack transparency, making them unsuitable for in-depth analysis and academic exploration. They also raise privacy and security concerns, especially when handling RTL designs that may contain sensitive intellectual property. In contrast, lightweight models that can run locally offer a secure, privacy-preserving alternative—enabling hardware engineers to integrate AI directly into their design workflows. However, existing open-source models still underperform on RTL tasks, largely due to the absence of high-quality, functionally validated RTL datasets in their training corpora~\cite{li2023starcoder,lozhkov2024starcoder}. While training algorithms are readily available, progress is bottlenecked by the lack of open datasets with functional correctness validation.

A key challenge in building such datasets lies in constructing large-scale, high-quality training data that pairs natural language specifications with RTL implementations. Despite efforts to mine RTL code from open-source repositories~\cite{dehaerne2023deep, pei2024betterv, thakur2023benchmarking, thakur2024verigen}, much of the collected data lacks validation and may not align with its intended functionality. To address this, recent work has turned to LLMs-either prompting them to synthesize RTL designs from keyword-based specifications~\cite{liu2024rtlcoder, liu2024rtlcoder2} or leveraging them to rewrite existing RTL code and generate matching specifications~\cite{cui2024origen, pei2024betterv, thakur2024verigen}. In both cases, syntax checkers are often employed to filter uncompilable code or provide feedback for iterative refinement, but these techniques still fall short of validating functional correctness.

As far as we know, all these prior work~\cite{liu2024rtlcoder,liu2024rtlcoder2,cui2024origen,pei2024betterv,thakur2024verigen} have focused solely on ensuring \emph{syntactic correctness}, overlooking \emph{functional correctness}. As a result, many dataset examples compile successfully but may not implement the behavior described in their natural language specifications. The distinction between syntactic correctness and functional correctness has important implications for model evaluation and real-world deployment. While functionally correct code inherently satisfies syntax constraints, syntactic correctness alone does not guarantee correct functionality. This gap is evident in the results reported by the RTLLM benchmark~\cite{lu2024rtllm}, where GPT-4o attains a high syntax accuracy of 100.0\%, yet achieve only 69.0\% in terms of functional correctness. Ultimately, in real-world settings, it is functional correctness rather than syntactic validity that truly matters.

\renewcommand\tabularxcolumn[1]{m{#1}}
\newcolumntype{L}[1]{>{\raggedright\arraybackslash}m{#1}}  
\newcolumntype{C}[1]{>{\centering\arraybackslash}m{#1}}    
\newcolumntype{Z}{>{\raggedright\arraybackslash}X}         

\newcommand{\yes}{\Large\textcolor{green!60!black}{\ding{51}}} 
\newcommand{\no }{\Large\textcolor{red!80!black}{\ding{55}}}

\begin{table*}[!tb]
  \small
  \centering
  \renewcommand{\arraystretch}{1.3}
  %
  \begin{tabularx}{\textwidth}{L{2.3cm}|L{3.5cm}|Z|C{1cm}|C{1cm}}
    \toprule
    \textbf{Prior Work}
      & \textbf{Strategy}
      & \textbf{Description}
      & \textbf{Syntax Checker}
      & \textbf{Unit Tests} \\
    \midrule

    RTLCoder~\cite{liu2024rtlcoder2}
      & Keyword-based Generation, Mutation
      & Prompt LLM with keywords and existing code, followed by iterative mutation to get instruction-code pairs.
      & \yes
      & \no \\
    \hline

    OriGen~\cite{cui2024origen}
      & Code‑to‑Code, Syntax Error Correction
      & Applies LLM‑driven code‑to‑code pipeline on existing RTL code and filters them by compiler error feedback. 
      & \yes
      & \no \\
    \hline


    BetterV~\cite{pei2024betterv}
      & Web Scraping \& Cleaning, Alignment with C
      & Large‑scale web‑collected Verilog, cleaned and filtered to enforce coding standards; aligns C with Verilog.
      & \yes
      & \no \\
    \hline

    VeriGen~\cite{thakur2024verigen}
      & Manually Collect Textbook and Open-Source Code
      & Mines real‑world RTL from GitHub and textbooks, manually cleans and organizes them into a structured dataset.
      & \yes
      & \no \\
    \hline

    ChipGPT~\cite{chang2024data}
      & AST‑based Synthesis
      & Converts Verilog ASTs into natural‑language prompts and injects semantic error variants via EDA‑tool feedback. 
      & \yes
      & \no \\
    \midrule

    \shortstack[l]{\textbf{\name{}}\\\textbf{(Our Work)}}
      & Feedback‑Directed Refinement, Simulation, Unit Test Generation
      & Iteratively generate unit tests with a teacher LLM, check implementations via compiler and simulator, and refining designs and tests until each design passes.
      & \yes 
      & \yes \\
    \bottomrule
  \end{tabularx}

  \caption{Comparison of Verilog fine-tuning dataset construction approaches.}
  \label{tab:related}
\end{table*}

In this work, we introduce \name{}, a model for RTL code generation fine-tuned on a high-quality dataset consisting of 125,777 examples that has been validated for functional correctness\footnote{While functional correctness is not fully guaranteed, we manually reviewed 100 randomly sampled examples and found that 92\% of the generated RTL code correctly matches the corresponding natural language descriptions.}. To construct this dataset, we develop a novel pipeline that combines unit test generation with feedback-directed refinement guided by a teacher LLM (GPT-4o-mini). Given a natural language specification and an initial RTL implementation, the teacher model first generates a unit test. If the RTL code fails the simulation, the model iteratively revises the design based on the observed error messages. When needed, the unit test is also updated to better reflect the intended functionality described by the specification. This process continues until the design passes simulation or a retry limit is reached. The resulting fine-tuning dataset consists of 125,777 validated triples: a natural language specification, a correct RTL design, and a self-checking unit test.

We fine-tune \name{} from \basemodel using our curated dataset and evaluate it on two established RTL code generation benchmarks: VerilogEval~\cite{liu2023verilogeval} and RTLLM~\cite{lu2024rtllm}. \name{} achieves new state‑of‑the‑art performance, achieving up to 71.7\% and 27.4\% relative gains in the pass@k metric over the previous best fine‑tuned model OriGen~\cite{cui2024origen}.

We conduct an ablation study demonstrating that models trained on our functionally validated dataset outperform those trained on non-validated data, under the same base model and training setup. These results highlight the importance of high-quality, functionally validated datasets for RTL code generation.

Our contributions are as follows:
\begin{itemize}
    \item We introduce \name{}, an RTL code generation model fine‑tuned on a dataset validated for functional correctness. On the VerilogEval and RTLLM benchmarks, \name{} achieves state‑of‑the‑art performance among open‑source fine‑tuned models, yielding relative pass@k gains of up to 71.7\% and 27.4\% over the prior best.

    \item We develop a dataset augmentation pipeline that combines unit test generation with feedback-directed refinement guided by a teacher LLM. This yields, to the best of our knowledge, the largest fine-tuning dataset to date with functional validation, consisting of 125,777 validated triples of natural language specifications, RTL designs, and passing tests.

    \item We conduct an ablation study showing that functional validation during dataset construction improves model performance, underscoring the importance of using high-quality functionally validated datasets for RTL code generation.
\end{itemize}

\begin{figure*}[!tb]
    \centering
    \includegraphics[width=\linewidth]{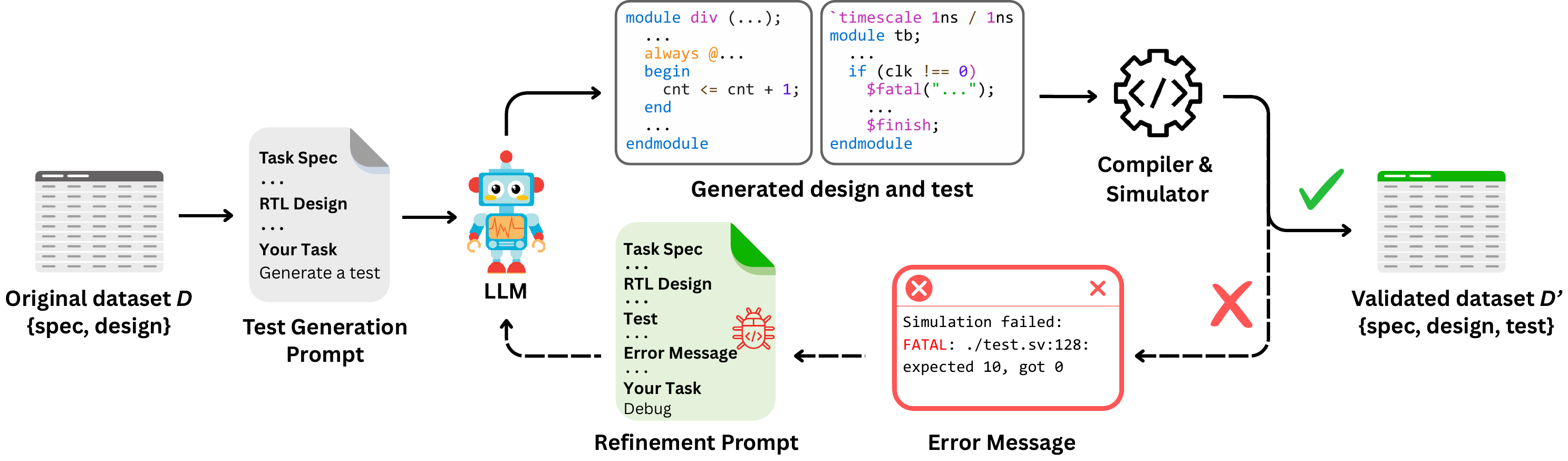}
    \caption{LLM-guided dataset augmentation overview.}
    \label{fig:overview}
\end{figure*}

\section{Background and Related Work}
\label{sec:background}

\subsection{Language Modeling and Fine-Tuning}
\label{subsec:llm}

Large Language Models (LLMs) are deep neural networks trained to perform \emph{language modeling}, a task where the model learns to predict the next token in a sequence. Formally, given a sequence of tokens \( x = (x_1, x_2, \ldots, x_T) \), the training objective is to maximize the log-likelihood:

\begin{equation}
\mathcal{L}_{\text{LM}} = \sum_{t=1}^{T} \log P(x_t \mid x_{<t}; \theta),
\end{equation}

where \( \theta \) denotes the model parameters and \( x_{<t} = (x_1, \ldots, x_{t-1}) \) represents the context tokens. This autoregressive objective enables the model to generate coherent text and capture long-range dependencies across various domains.

The training of LLMs is typically organized into two stages:

\begin{itemize}
    \item Pre-training: The model is trained on massive, diverse corpora (e.g., web data, books, source code) to acquire broad knowledge and language understanding. This stage is expensive and performed once per model.
    \item Post-training: The pre-trained model is adapted to specific tasks using smaller, curated datasets. This stage includes supervised fine-tuning (SFT), where the model is trained on task-specific input-output pairs.
\end{itemize}

Since post-training large models from scratch is resource-intensive, researchers have developed parameter-efficient fine-tuning methods. One widely used approach is \emph{Low-Rank Adaptation (LoRA)}~\cite{hu2022lora}. Instead of updating the full weight matrices \( W \in \mathbb{R}^{d \times k} \) in each linear layer, LoRA freezes the original weights and introduces a trainable low-rank update:

\begin{equation}
W' = W + \Delta W = W + A B,
\end{equation}

where \( A \in \mathbb{R}^{d \times r} \) and \( B \in \mathbb{R}^{r \times k} \), and \( r \ll \min(d, k) \). Only \( A \) and \( B \) are updated during training, while \( W \) remains unchanged. This technique reduces both memory and compute overhead during adaptation, making it feasible to specialize large LLMs to domain-specific applications, such as RTL generation, with limited computational resources.

\begin{figure*}[!tb]
  \centering
  \captionsetup[subfigure]{justification=centering,singlelinecheck=false}

  \begin{subfigure}[t]{0.35\textwidth}
    \vspace{0pt} 
    \centering
    \begin{promptbox}
Create a Verilog module \texttt{and3} with three one-bit inputs \texttt{a}, \texttt{b}, \texttt{c}, and a one-bit output \texttt{y}. The output should perform a bitwise AND across all three inputs using a procedural block. The truth table:

\medskip
\begin{center}
\begin{tabular}{cccc}
\toprule
$a$ & $b$ & $c$ & $y$ \\\midrule
0 & x & x & 0 \\
x & 0 & x & 0 \\
x & x & 0 & 0 \\
1 & 1 & 1 & 1 \\\bottomrule
\end{tabular}
\end{center}
    \end{promptbox}
    \vspace{2.5pt}
    \caption{Natural language specification taken from the Origen~\cite{cui2024origen} dataset.}
    \label{fig:spec}
  \end{subfigure}
  \hfill
  \begin{subfigure}[t]{0.3\textwidth}
    \vspace{0pt}
\begin{lstlisting}[style=verilog-style]
module and3(
    input  wire a,
    input  wire b,
    input  wire c,
    output reg  y
);

always @* begin
    y <= a;
    y <= y & c;
    y <= y & b;
end

endmodule
\end{lstlisting}
    \caption{Buggy design taken from the Origen~\cite{cui2024origen} dataset. It times out on the generated test shown in \Cref{fig:test}.}
    \label{fig:buggy}
  \end{subfigure}
  \hfill
  \begin{subfigure}[t]{0.3\textwidth}
    \vspace{0pt}
\begin{lstlisting}[style=verilog-style]
module and3(
    input  wire a,
    input  wire b,
    input  wire c,
    output reg  y
);


always @* begin
    // combinational AND
    y = a & b & c;
end

endmodule
\end{lstlisting}
    \caption{Correct design fixed by the teacher model that passes the generated test in \Cref{fig:test}.}
    \label{fig:fixed}
  \end{subfigure}

  \caption{Natural language specification (left) and the corresponding buggy and corrected Verilog designs (middle and right). The specification and buggy design are from the original dataset~\cite{cui2024origen}, which lacks tests, while the test (\Cref{fig:test}) and corrected design are generated by a teacher model (GPT-4o-mini) and included in our validated dataset.}
  \label{fig:motivating}
\end{figure*}

\begin{figure}[!tb]
  \centering
\begin{lstlisting}[style=verilog-style]
`timescale 1ns/1ps
module tb_and3;
  reg  a = 0, b = 0, c = 0;
  wire y;

  // Instantiate the DUT (Design Under Test)
  and3 uut (.a(a), .b(b), .c(c), .y(y));

  initial begin
    // Wait for signals to settle
    #1;

    // Set all inputs to 1; expected y = 1
    {a, b, c} = 3'b111;
    #1;

    // Check output, report error if incorrect
    if (y !== 1'b1)
      $fatal(1, "FAIL: y=%b (expected 1)", y);
    
    $display("PASS");
    $finish;
  end
endmodule
\end{lstlisting}
\caption{Unit test for the \texttt{and3} module. The buggy design (\Cref{fig:buggy}) times out on this test, while the corrected design (\Cref{fig:fixed}) passes successfully. The test is generated by the teacher model GPT-4o-mini using the prompt in \Cref{fig:testprompt}, and is used to validate and augment the original dataset, which contains no tests.}
\label{fig:test}
\end{figure}

\subsection{Related Work on RTL Code Generation}
\label{subsec:dataset}

Progress on open-source RTL code generation is limited by the absence of large-scale, high-quality datasets. To mitigate this, recent efforts have focused on automated data mining and augmentation techniques to enrich existing corpora of RTL examples. \Cref{tab:related} presents the comparison of different strategies for constructing fine-tuning datasets.

Mining open-source RTL designs is a common strategy for dataset construction. VeriGen~\cite{thakur2024verigen} compiles Verilog modules from GitHub and textbooks into a structured corpus using automated syntax checks. BetterV~\cite{pei2024betterv} collects Verilog modules from the internet and then filters designs based on coding style and syntactic validity. CraftRTL~\cite{Liu2024CraftRTLHS} augments fine-tuning data with non-textual code representations, injecting synthetic errors derived from intermediate model checkpoints into open-source Verilog code. Other works~\cite{cui2024origen,zhang2024mg,goh2024english} adopt similar methodologies for sourcing and preprocessing RTL code.

Another line of work leverages a commercial LLM for synthetic data generation. RTLCoder~\cite{liu2024rtlcoder} prompts GPT‑3.5 with domain keywords to generate both task descriptions and corresponding RTL, discarding any outputs that fail to compile. OriGen~\cite{cui2024origen} further employs Claude 3.5 in a two‑stage code‑to‑code pipeline: first turning mined RTL code into natural language specifications, then regenerating code from these specifications under compiler guidance, combining the strengths of real-world examples and synthetic generation. ChipGPT~\cite{chang2024data} transforms Verilog ASTs into natural language specifications.

While most of the existing work listed in Table~\ref{tab:related} ensures syntax validity, none of them has any evidence of functional correctness. Without comprehensive unit tests or simulation‑based feedback during dataset construction, models fine‑tuned on these corpora may produce code that compiles but still fails to meet the intended natural language specification.

A recent work, OpenLLM-RTL~\cite{liu2024openllm}, explores the idea of using LLMs to generate assertions, producing a functionally verified dataset of 7k examples. While sharing the same goal of improving functional correctness in fine-tuning datasets, our work takes a different approach by generating unit tests for validation. Our final dataset contains over 125,777 examples—the largest functionally validated RTL dataset to date.

Beyond data collection and synthesis techniques, several works explore other methods to enhance RTL code generation quality. ScaleRTL~\cite{Deng2025ScaleRTLSL} emphasizes reasoning by generating intermediate traces and leveraging test-time compute through iterative self-reflection. DeepRTL~\cite{Liu2025DeepRTLBV} adopts curriculum learning guided by multi-level natural language summaries. VeriSeek~\cite{Wang2024LargeLM} applies reinforcement learning with feedback derived from AST-level similarity between LLM outputs and reference designs. AutoVCoder~\cite{Gao2024AutoVCoderAS} incorporates retrieval-augmented generation (RAG), dynamically supplying relevant Verilog snippets to the model. CodeV~\cite{Zhao2024CodeVEL} extends generation capabilities to tasks such as fill-in-the-middle (FIM). Our work adopts standard supervised fine-tuning while focusing on constructing a large-scale, functionally validated dataset. Our approach is complementary and orthogonal to existing techniques.

\section{Methodology}
\label{sec:method}

\subsection{Overview}
\label{subsec:overview}

We aim to improve the quality of fine-tuning datasets consisting of natural language specifications paired with syntactically correct Verilog designs, as seen in prior work~\cite{liu2024rtlcoder,liu2024rtlcoder2,cui2024origen,pei2024betterv,thakur2024verigen}. These datasets, including Origen~\cite{cui2024origen}, contain Verilog designs that pass syntax checks but are not validated against unit tests to ensure functional correctness. To address this limitation, we introduce an automated dataset augmentation pipeline that leverages a teacher language model, e.g., GPT-4o-mini, to validate each example through iterative refinement.
As illustrated in \Cref{fig:overview}, given a natural language specification and an initial RTL design, the teacher model first generates a unit test. If the RTL design fails the simulation, the model iteratively revises the design based on the error message. When needed, it also updates the unit test to better align with the natural language specification. Although our experiments focus on augmenting the Origen dataset due to its size and quality, the proposed methodology is broadly applicable to any dataset lacking test validation.

The pipeline begins with the original dataset $D = \{(\text{specification}, \text{design})\}$, where each RTL design is intended to implement a corresponding natural language specification. However, because no tests are provided, there is no evidence that the designs exhibit the intended functional behavior. For each pair, we prompt the teacher model, GPT-4o-mini, to generate a unit test for the design. The test is compiled and simulated with the design to check for correctness, where correctness means the design passes the simulation test.

If the simulation fails, we extract the resulting error message and re-invoke the teacher model using a refinement prompt. This prompt includes the specification, the current design and test, and the error message. The model attempts to resolve the failure by making minimal modifications to the design, the test, or both. This refinement process repeats iteratively: each candidate is re-simulated, and the cycle continues until the design passes the test or a maximum number of attempts is reached.

The final output is a validated dataset $D' = \{(\text{specification}, \text{design}, \text{test})\}$, where each triplet contains a natural language specification, a Verilog design, and unit tests. A concrete motivating example is shown in \Cref{subsec:example}, and the details of the algorithm and prompts are provided in \Cref{subsec:algorithm}.

\begin{figure*}[tb]
  \centering
  \captionsetup[subfigure]{justification=centering,singlelinecheck=false}

  \begin{subfigure}[t]{0.48\textwidth}
    \centering
    \begin{templatebox}
    \textbf{System Prompt}  
    You are a Verilog design and testing expert.  
    Given a hardware specification described in natural language, your job is to generate both a correct Verilog module and a corresponding unit test that checks its functionality through simulation.
    
    \medskip
    \textbf{User Prompt}  
    \begin{itemize}[nosep,leftmargin=*]
      \item \textit{Natural Language Specification}: \{NL Spec\}
      \item \textit{Initial Implementation}: \{design\}
      \item Your task:
       \begin{enumerate}[nosep]
          \item Provide the unit tests for the given design.
          \item Revise the Verilog implementation if the original design fails to pass your test cases.
          \item Follow good coding practices, such as using meaningful comments to document key logic and decision points.
          \item Use \verb|$fatal(1, "msg")| to flag incorrect behavior.
          \item Output format: \{\texttt{"design": "...", "test": "..."}\}
        \end{enumerate}
    \end{itemize}
    \end{templatebox}
    \caption{Prompt for generating a Verilog module's corresponding test}
    \label{fig:testprompt}
  \end{subfigure}
  \hfill
  \begin{subfigure}[t]{0.48\textwidth}
    \centering
    \begin{templatebox}
    \textbf{System Prompt}  
    You are a Verilog design and testing expert.  
    Analyze a failing design and its test, and make minimal yet sufficient edits to correct the issue while preserving the intended behavior specified in natural language.
    
    \medskip
    \textbf{User Prompt}  
    \begin{itemize}[nosep,leftmargin=*]
      \item \textit{Natural Language Specification}: \{NL Spec\}
      \item \textit{Previous Design and Test}: \{design\}, \{test\}
      \item \textit{Simulation Output}: \{error message\}
      \item Your task:
        \begin{enumerate}[nosep]
          \item Carefully identify the root cause of the failure by analyzing the code and the error message.
          \item Make changes to either the design or the test (or both) to resolve the issue while maintaining correctness.
          \item Output format: \{\texttt{"explanation": "...", "design": "...", "test": "..."}\}
        \end{enumerate}
    \end{itemize}
    \end{templatebox}
    \caption{Prompt for refining a failing Verilog design and test}
    \label{fig:refineprompt}
  \end{subfigure}

  \caption{Prompt templates provided to the teacher model for automated Verilog test generation and refinement, ensuring that the final design passes the generated test and matches the original natural language specification.}
  \label{fig:prompts}
\end{figure*}

\subsection{Motivating Example}
\label{subsec:example}

\Cref{fig:motivating} presents a motivating example taken directly from the Origen dataset~\cite{cui2024origen}, highlighting a key limitation of datasets that rely only on syntax checks for validation. Prior work in RTL generation typically assumes that syntactic correctness is sufficient for fine-tuning, without verifying functionality through unit tests. This example demonstrates that a design can compile without errors yet fail to implement the intended behavior. It also illustrates how our method can automatically detect and correct such issues through test generation and iterative refinement.

This example includes a natural language specification (\Cref{fig:spec}), a buggy RTL design from the original dataset (\Cref{fig:buggy}), and a corrected design produced by our pipeline (\Cref{fig:fixed}). The specification describes a simple combinational module, \texttt{and3}, which computes the bitwise AND of three one-bit inputs: \texttt{a}, \texttt{b}, and \texttt{c}.

The original design, though syntactically valid, is functionally incorrect due to several semantic issues. First, it misuses non-blocking assignments (\texttt{<=}) inside a combinational \texttt{always @*} block, which can lead to counterintuitive synthesis results. Second, if instead used inside a sequential block, the sequence of non-blocking assignments in the design---\texttt{y <= a}, then \texttt{y <= y \& c}, and finally \texttt{y <= y \& b}---does not correctly compute and store in \texttt{y} the bitwise AND of \texttt{a}, \texttt{b}, and \texttt{c}. In particular, non-blocking assignments defer updates until the end of the current timestep, meaning that all assignments operate on the same initial value of \texttt{y}, and only the final assignment takes effect. Finally, if the non-blocking assignments were replaced with blocking ones, the code would introduce a combinational feedback loop, which cannot stabilize.

These types of errors occur because the RTL code in prior datasets, including Origen~\cite{cui2024origen}, is synthetically generated by teacher LLMs such as Claude 3.5 and filtered only through syntax checks. Without simulation or test-based validation, semantic bugs that affect functional correctness remain undetected.

We provide the natural language specification and the buggy RTL design to the teacher model GPT-4o-mini, prompting it to generate a unit test using the template shown in \Cref{fig:testprompt} (further detailed in \Cref{subsec:algorithm}). The resulting test is shown in \Cref{fig:test}, which sets all three inputs to \texttt{1} and checks whether the output \texttt{y} evaluates to \texttt{1} as expected. When the buggy design (\Cref{fig:buggy}) is simulated with this test, it hangs and ultimately times out. The bug exemplifies a combinational loop. The always @* block is meant for combinational logic and its evaluation is triggered upon changes to any of the variables read inside the block. In this case, an evaluation of the block is triggered when either y, a, b, or c changes. However, y is both read (on the RHS) and written (on the LHS) in the same block. Upon evaluating the block, it schedules an update to y, which causes a change to y. This change retriggers the block, leading to another scheduled update to y, and so on. This loop continues indefinitely, preventing the simulation from converging.

The corrected version replaces the non-blocking assignments with a single blocking assignment (\texttt{=}), ensuring that \texttt{y} is updated immediately with the result of \texttt{a \& b \& c}, as required by the specification. This version passes the test generated by the teacher model and behaves correctly under simulation.

This example underscores the importance of functional validation in RTL datasets. Syntax checks alone cannot catch subtle but critical semantic errors. Our methodology, through teacher-driven test generation and iterative refinement, ensures that each design in the augmented dataset is not only syntactically valid but also functionally validated with unit tests.

\subsection{Algorithm and Prompts}
\label{subsec:algorithm}

\algrenewcommand\algorithmicrequire{\textbf{Input:}}
\algrenewcommand\algorithmicensure{\textbf{Output:}}

\algrenewcommand\algorithmicforall{\textbf{for each}}

\begin{algorithm}[!tb]
\caption{Dataset Augmentation with a Teacher LLM}
\label{alg:refine}
\begin{algorithmic}[1]
  \Require Original dataset 
    $D = \{(s_i, d_i)\}_{i=1}^N$
  \Statex \hspace{\algorithmicindent}\Comment{$s_i$: NL specification; $d_i$: RTL design}
  \Statex \hspace{\algorithmicindent}Maximum attempts $T$
  
  \Statex \hspace*{-16.5pt}\textbf{Define:} $\mathit{GenTestTpl}\gets$ prompt template for test generation 
  \Statex \hspace{\algorithmicindent}$\mathit{RefineTpl}\gets$ prompt template for iterative refinement
  
  \Ensure Augmented dataset 
    $D' = \{(s_i,d_i,t_i)\}_{i=1}^M$
  \Statex \hspace{\algorithmicindent}\Comment{$t_i$: Generated unit test}

  \State $D' \gets \varnothing$
  \For{each $(s,d)\in D$}
    \State $attempt\gets 0,\; success\gets \mathrm{false}$
    \While{$attempt < T \,\wedge\, \neg success$}
      \State $attempt\gets attempt + 1$
      \If{$attempt == 1$}
        \State $d,t \gets \mathrm{LLMInvoke}(\mathit{GenTestTpl},s,d)$
      \Else
        \State $d,t\gets \mathrm{LLMInvoke}(\mathit{RefineTpl},s,d,t,err)$
      \EndIf
      \State $success, err\gets \mathrm{RunVerilogTest}(d,t)$
      \If{$success$}
        \State $D'\gets D'\,\cup\,\{(s,d,t)\}$
      \EndIf
    \EndWhile
  \EndFor
  \State \Return $D'$
\end{algorithmic}
\end{algorithm}

\Cref{alg:refine} presents our automated pipeline for transforming an unvalidated RTL dataset into a functionally validated one. Starting from a dataset $D = \{(s_i, d_i)\}_{i=1}^N$, where each example consists of a natural language specification $s_i$ and a corresponding RTL design $d_i$ (e.g., from Origen~\cite{cui2024origen}), the goal is to generate a unit test $t_i$ that validates the functional correctness of the design. If the design fails to pass the test, we invoke an iterative refinement loop that updates the design and test until it passes or a maximum number of attempts $T$ is reached. We set $T = 5$ in our experiments.

The procedure is powered by a teacher model, GPT-4o-mini, which corresponds to the \texttt{LLMInvoke} calls in \Cref{alg:refine}. While stronger models such as GPT-4o or o3-mini may yield better performance, we use GPT-4o-mini in practice because of the large size of the dataset (217{,}462 examples in Origen) and the high cost associated with repeated API queries to OpenAI models.

The process begins by prompting the teacher model with the test generation template (\Cref{fig:testprompt}), together with a natural language specification and its initial RTL design (e.g., \Cref{fig:spec} and \Cref{fig:buggy}). The model then produces a candidate unit test (e.g., \Cref{fig:test}) designed to check whether the design satisfies the intended functionality under simulation.

The design and test are compiled and simulated using standard Verilog tooling. If the test fails, for example due to a timeout, incorrect output, or another runtime error, we construct a refinement prompt (\Cref{fig:refineprompt}) that includes the specification, the failing design and test, and the simulation error message (corresponding to the \texttt{err} variable in \Cref{alg:refine}). This prompt is then passed to the teacher model, which attempts to fix the issue by making edits to the design, the test, or both.

The refinement process repeats until the updated design passes simulation or the maximum number of attempts $T$ is reached. Once a design successfully passes its test, the validated triple $(s_i, d_i, t_i)$ is added to the output dataset $D'$.

This strategy enables systematic detection and correction of subtle RTL bugs that cannot be identified through syntax checks alone. By integrating LLM-based test generation and iterative refinement into the dataset construction pipeline, we produce a dataset that is not only syntactically valid but also functionally validated through simulation.

While functional correctness under all possible inputs cannot be guaranteed, the inclusion of unit tests makes our augmented dataset substantially more robust than prior approaches that rely solely on syntactic checking. We view this as a practical and scalable step toward building higher-quality fine-tuning datasets for RTL generation. To assess quality, we manually reviewed 100 randomly sampled examples and found that 92\% of the generated RTL code correctly matched the corresponding natural language descriptions.
\section{Experimental Setup}
\label{sec:setup}

\subsection{Dataset}

Following the methodology described in Section~\ref{sec:method}, we construct a fine-tuning dataset comprising 125,777 examples. Each example includes a natural language specification, a corresponding RTL design, and associated unit tests. Table~\ref{tab:dataset} summarizes key statistics: the specifications contain an average of 247 words (ranging from 116 to 549), RTL implementations average 35 lines of code (ranging from 5 to 225), and unit tests average 55 lines (ranging from 6 to 197). We use the specification–solution pairs from this dataset to train our model, \name{}.

\begin{table}[!tb]
\centering
\small
\begin{tabular}{lcccc}
\toprule
\multirow{2}{*}{Category} & \multirow{2}{*}{Count} & \multicolumn{3}{c}{Length} \\
\cmidrule(lr){3-5}
                          &                             & Min & Max & Avg \\
\midrule
NL specification (words)        & \multirow{3}{*}{125,777} & 116 & 549 & 247 \\
Design (lines of RTL)  & &  5 & 225 & 35 \\
Unit tests (lines of RTL) & &  6 & 197 & 55 \\
\bottomrule
\end{tabular}
\caption{Dataset statistics: total number of examples and length distributions for natural language specifications, RTL implementations, and unit tests in the \name{} dataset.}
\label{tab:dataset}
\end{table}

\subsection{LoRA Fine‑Tuning Setup}
\label{subsec:lorasetup}

Following standard practices for LLM fine-tuning, we fine-tune the base model of \basemodel{} using Low-Rank Adaptation (LoRA, described in \Cref{subsec:llm}), with a rank of 16 and a scaling factor of 32 to all linear projection layers in the transformer. Training is conducted over 3 epochs with a batch size of 40. We adopt a constant learning rate of $1 \times 10^{-5}$, paired with a linear decay scheduler and a warm-up ratio of 0.05. The optimizer is used with a weight decay of $1 \times 10^{-4}$, and gradient clipping is applied with a maximum norm of 1.

\begin{table*}[!tb]
\centering
\small
\begin{tabular}{ll|ccc|ccc|c|c}
\toprule
\multirow{4}{*}{\textbf{Model Type}} & \multirow{4}{*}{\textbf{Evaluated Model}} & 
\multicolumn{6}{c|}{\textbf{VerilogEval V1.0~\cite{liu2023verilogeval}}} & 
\multicolumn{2}{c}{\textbf{RTLLM V1.1~\cite{lu2024rtllm}}} \\
& & 
\multicolumn{6}{c|}{(using pass@k metric)} & 
\multicolumn{2}{c}{(using pass@5 metric)} \\
\cmidrule(lr){3-8} \cmidrule(lr){9-10}
& & 
\multicolumn{3}{c|}{Eval-Machine (\%)} & 
\multicolumn{3}{c|}{Eval-Human (\%)} & 
\multirow{2}{*}{\begin{tabular}{@{}c@{}}Syntax-VCS\\(\%)\end{tabular}} & 
\multirow{2}{*}{\begin{tabular}{@{}c@{}}Functional\\(\%)\end{tabular}} \\
& & k=1 & k=5 & k=10 & k=1 & k=5 & k=10 & & \\
\midrule

\multirow{10}{*}{\textbf{Base Models}}
& o4-mini-2025-04-16 & 61.9	& 67.8	& 68.6	& \textbf{64.3}	& 66.4	& 67.1	& 86.2	& \textbf{72.4} \\
& GPT-4o-2024-11-20 & 63.7 & 66.5 & 67.1 & 54.3 & 60.4 & 62.2 & \textbf{100.0} & 69.0 \\
& GPT-4o-mini-2024-07-18 & 55.7 & 62.4 &	64.3	& 44.7 &	51.6 &	55.1 & 89.7	& 65.5 \\
& DeepSeek-R1 & 65.7 & 70.9 & \textbf{72.0} & 62.8 & \textbf{69.1} & 69.9 & 79.3 & 58.6 \\
& o3-mini-2025-01-31 & \textbf{66.4} & \textbf{71.6} & \textbf{72.0} & 62.0 & 68.9 & \textbf{69.9} & 69.0 & 55.2 \\
& Qwen2.5-14B-Instruct & 47.8 & 54.2 & 55.2 & 35.3 & 40.0 & 42.3 & 69.0 & 41.4 \\
& Gemini-2.0-flash-001 & 60.3 & 62.6 & 63.6 & 52.1 & 57.6 & 59.0 & 65.5 & 34.5 \\
& DeepSeek-R1-Distill-Qwen-14B & 46.2 & 64.1 & 68.5 & 36.7 & 51.7 & 55.1 & 62.1 & 34.5 \\
& DeepSeek-Coder-7B-v1.5 & 44.4	& 58.9 &	62.9 &	25.8 &	40.2 &	44.9 &	48.3 &	24.1 \\
& LLaMA-2-7B & 7.0	& 15.6	& 18.9	& 0.4	& 2.1	& 3.8 &	3.4	& 0.0 \\

\midrule
\multirow{5}{*}{\shortstack[l]{\textbf{Fine-Tuned Models}\\\textbf{(Prior Work)}}}
& OriGen~\cite{cui2024origen} & 35.9 &	\cellcolor{gray!25}65.1 &	\cellcolor{gray!25}68.5 &	22.3 &	47.5 &	51.9 &	51.7 &	37.9 \\
& RTLCoder-DeepSeek~\cite{liu2024rtlcoder} & 22.0 &	51.4 &	57.3 & 14.7 &	35.2 &	42.3 &	17.2 &	10.3 \\
& RTLCoder-Mistral~\cite{liu2024rtlcoder} & 17.6 &	46.4 &	56.6 &	12.4 &	31.5 &	36.5 &	3.4 &	0.0 \\
& ChipGPT-LLaMA3.1-8B-SFT~\cite{chang2024data} & 17.6 &	46.4 &	56.6 &	12.4 &	31.5 &	36.5 &	13.8 &	0.0 \\
& ChipGPT-LLaMA2-SFT-7B~\cite{chang2024data} & 0.9 &	4.2 &	7.7 &	0.6 &	2.2 &	3.8 &	6.9 &	0.0 \\

\midrule
\multirow{1}{*}{\textbf{Our Work}} 
& \name & \cellcolor{gray!25}55.7 &	62.9 &	64.3 &	\cellcolor{gray!25}38.3 &	\cellcolor{gray!25}49.2 &	\cellcolor{gray!25}51.9 &	\cellcolor{gray!25}79.3 &	\cellcolor{gray!25}48.3 \\

\bottomrule
\end{tabular}
\caption{RTL code generation performance across models. To ensure a fair comparison, we use the same input prompts and apply identical post-processing scripts, running inference with model weights released by prior work.}
\label{tab:maineval}
\end{table*}

\subsection{Benchmarks and Metrics}

Following the evaluation protocol established in prior work~\cite{liu2024rtlcoder2,cui2024origen}, we benchmark against VerilogEval~\cite{liu2023verilogeval} and RTLLM~\cite{lu2024rtllm}. For VerilogEval, we report the standard \textit{Pass@k} metric with $k \in \{1, 5, 10\}$, which estimates the expected probability that at least one of the top-$k$ generated programs passes all test cases. The metric is defined as:
\[
\text{Pass@}k = \mathbb{E}\left[1 - \frac{\binom{n - c}{k}}{\binom{n}{k}}\right]
\]
where $n$ is the total number of generated programs and $c$ is the number of correct ones. All test cases are manually created by experts who design the benchmarks. In all evaluations, we set $n = 10$. For RTLLM, we report both syntax correctness and functional correctness using Pass@5. This evaluation setup aligns with that used in prior work~\cite{cui2024origen}.

\subsection{Models for Evaluation}
We evaluate two groups of models. The first group consists of pretrained-only base models, including OpenAI’s latest releases (o4-mini, o3-mini, GPT-4o, GPT-4o-mini), Google’s Gemini 2.0 Flash, DeepSeek’s R1 and DeepSeek-Coder-7B-v1.5 (the base model used in prior work~\cite{cui2024origen}), Meta’s LLaMA2-7B model, and Alibaba’s \basemodel (our base model for fine-tuning). The second group includes fine-tuned models with released weights from prior work: Origen~\cite{cui2024origen}, RTLCoder~\cite{liu2024rtlcoder}, and ChipGPT~\cite{chang2024data}.

To ensure a fair comparison, we use identical input prompts and post-processing scripts across all models. For models released by prior work, we do not adopt their model-specific prompts~\cite{cui2024origen} or inference pipelines~\cite{chang2024data,liu2024rtlcoder}. Instead, we apply a uniform evaluation script, with the only variable being the model under test. This standardization is critical, as both input formatting and post-processing can significantly affect performance. By controlling these factors, we isolate model capability and enable a fair comparison.
\section{Results}
\label{sec:results}

\subsection{Main Evaluation Results}

\Cref{tab:maineval} shows the results. Our major findings are as follows:

\paragraph{Comparison with prior work}
\name{} achieves state-of-the-art results across two RTL code generation benchmarks, outperforming all previously released open-source fine-tuned models. On VerilogEval-Machine, \name{} attains a pass@1 accuracy of 55.7\%, representing a 19.8 percentage point improvement over the best prior model, OriGen. On VerilogEval-Human, it reaches 38.3\%, exceeding OriGen by 16.0 percentage points. Across all evaluated $k$-shot settings ($k{=}1,5,10$), \name{} consistently maintains its lead on the Human split. On the RTLLM benchmark, \name{} achieves 79.3\% syntax correctness and 48.3\% functional correctness, surpassing OriGen’s 51.7\% and 37.9\%, respectively. In conclusion, \name{} delivers relative improvements of up to 71.7\% on VerilogEval and 27.4\% on RTLLM in pass@k accuracy, surpassing the previous state-of-the-art model on both benchmarks.

To better understand the relatively low performance of ChipGPT~\cite{chang2024data}, we examined its outputs in detail. We observed that its generated RTL designs often include module headers that deviate from the given specifications, revealing difficulty in precise instruction following. Moreover, its base model, LLaMA2-7B, performs even worse, suggesting that limitations in the instruction-following capabilities of the underlying pretrained model constrain the effectiveness of the fine-tuned variant. For a fair comparison, we do not apply any of the model-specific customized post-processing scripts that attempt to fix syntax or header issues. Instead, we use a standardized evaluation script for all models, extracting Verilog code as-is to ensure consistency.

\paragraph{Effectiveness of our fine-tuning}
Starting from Qwen‑2.5‑14B‑Instruct as our base model, \name{} delivers substantial gains across VerilogEval. On the VerilogEval-Machine split, pass@1 jumps up by 7.6\%, pass@5 by 4.0\%, and pass@10 by 2.1\%, and VerilogEval-Human reflects the same trend. On RTLLM, functional pass@5 is 7\% higher than its base model. Specifically, \name{} even marginally outperforms one of the commercial models, Google's Gemini-2.0-flash, on pass@5 and pass@10 metrics of Eval-Machine as well as on RTLLM. Together, these results demonstrate that our fine‑tuning process and our validated dataset significantly boost pass@k metrics and semantic correctness in RTL generation.

\paragraph{Model gap remains}

Despite the observed improvements, a substantial performance gap persists between \name{} and the strongest large models. For instance, o3-mini attains 66.4\% on VerilogEval Pass@1 compared to \name{}'s 55.7\%. DeepSeek-R1 achieves 69.1\% on human-graded Pass@5, versus \name{}'s 49.2\%. Commercial LLMs such as GPT-4o reach a perfect 100.0\% Syntax-VCS validity and 69.0\% functional correctness, while \name{} records 79.3\% and 48.3\%, respectively. Despite the performance gap, open-source lightweight models offer compelling advantages. They provide transparency, allow for local deployment, and ensure intellectual property protection, i.e., capabilities that are particularly important for RTL design workflows where security, customizability, and integration into existing toolchains are critical.

\subsection{Ablation Study of Dataset}

\begin{table}[!tb]
\centering
\small
\begin{tabular}{lccc}
\toprule
\multirow{2}{*}{\textbf{Model}} &
  \multirow{2}{*}{\shortstack[c]{\textbf{VerilogEval}~\cite{liu2023verilogeval}\\(Pass@5)}} &
  \multicolumn{2}{c}{\shortstack[c]{\textbf{RTLLM}~\cite{lu2024rtllm}\\(Pass@5)}} \\
\cmidrule(lr){3-4}
& & Syntax & Func \\
\midrule
Qwen2.5-14B-Instruct (base)        & 46.8 & 69.0 & 41.4 \\
Qwen w/ unvalidated data & 53.5 & 75.9 & 44.8 \\
Qwen w/ validated data                   & 55.8 & 79.3 & 48.3 \\
\bottomrule
\end{tabular}
\caption{We performed fine-tuning on the same base model using a functionally validated dataset and the functionally unvalidated dataset~\cite{cui2024origen}. We report Pass@5 metrics for all models on two benchmarks.}
\label{tab:dataset_ablation}
\end{table}

To assess the impact of dataset quality on RTL code generation, we conduct an ablation study using the same base model, \basemodel, fine-tuned on two datasets: (1) the unvalidated OriGen dataset from prior work~\cite{cui2024origen}, and (2) our newly curated, functionally validated dataset. All factors, including dataset size, fine-tuning hyperparameters, training procedures, and evaluation settings, are held constant to ensure a fair comparison.

Across all metrics, we observe a consistent improvement as dataset quality increases. On the VerilogEval benchmark (covering both Machine and Human subsets), the base model achieves 46.8\% Pass@5. Fine-tuning on the unvalidated dataset raises performance to 53.5\%, while our validated dataset further improves it to 55.8\%. For RTLLM syntax correctness, the trend is similar: 69.0\% for the base model, 75.9\% for the unvalidated version, and 79.3\% when trained on validated data. Functional correctness sees even more significant improvement, rising from 41.4\% (base) to 44.8\% (unvalidated) and ultimately to 48.3\% (validated).

These results demonstrate that functionally validated data provides more effective supervision than existing unvalidated data. This also underscores the importance of dataset quality in fine-tuning LLMs for RTL code generation.

\subsection{Test Passing Rates of Non-Validated Datasets}

We examine the quality of fine-tuning datasets released by prior work by evaluating their passing rates against our synthetic unit tests generated by the teacher model GPT-4o-mini. For each corpus, we randomly sample 1,000 Verilog implementations and apply the test generation and refinement pipeline described in \Cref{sec:method}. We then run corresponding unit tests against the original design and measure the proportion of the original designs that successfully pass the generated tests. As shown in Table~\ref{tab:dataset_failure}, only 24.4\% examples of the RTLCoder dataset~\cite{liu2024rtlcoder} pass our functional tests, while OriGen~\cite{cui2024origen} reaches 53.5\%.

OriGen’s higher pass rate aligns with its stronger code generation results in \Cref{tab:maineval}, hinting at a positive link between dataset validity and downstream performance. These findings highlight the potential value of incorporating functional correctness validation into fine-tuning dataset curation for better RTL code generation.

\begin{table}[!tb]
\centering
\small
\begin{tabular}{l c c}
\toprule
\textbf{Prior Datasets} 
  & \textbf{\# Sampled Examples} 
  & \textbf{Test Passing} (\%) \\
\midrule
RTLCoder~\cite{liu2024rtlcoder} 
  & 1000 
  & 24.4 \\
OriGen~\cite{cui2024origen}     
  & 1000                
  & 53.5 \\
\bottomrule
\end{tabular}
\caption{Test passing rates (\%) of datasets released by prior work on a randomly sampled set of 1000 examples.}
\label{tab:dataset_failure}
\end{table}

\section{Discussion and Future Work}
\label{sec:discussion}

While VeriCoder, combining unit test generation with feedback-driven refinement, improves the functional correctness of generated RTL code, it does not fully guarantee correctness. Synthetic test cases may fail to capture all possible edge cases. To address this challenge, future work should explore integrating formal verification techniques into the dataset construction pipeline to rigorously ensure the correctness of the generated code. Recent advancements have demonstrated promising results in translating natural language instructions into formal specifications~\cite{10918520, cosler2023nl2spec}, as well as enforcing formal constraints during LLM-based code generation~\cite{aggarwal2024alphaverus}.

Moreover, most existing approaches, including VeriCoder, focus on small-scale RTL generation. However, practical hardware development often involves large, repository-level codebases with intricate cross-file dependencies and requirements for long-range context~\cite{jimenez2023swe, suresh2025cornstack, jain2024r2e}. Recent work has begun to address these challenges through techniques such as combining fine-tuning with retrieval-augmented RTL code generation~\cite{wu2025rtlrepocoder,li2025deepcircuitx}. Extending VeriCoder’s unit test generation and feedback-directed refinement components to the repository scale will enable LLMs to handle more real-world RTL tasks.

Furthermore, reinforcement learning (RL) offers a powerful framework for further optimizing large language models' performance beyond what is achievable through supervised fine-tuning alone. Recent studies have demonstrated the effectiveness of RL in enhancing LLM-based code generation by incorporating diverse forms of feedback, such as test case outcomes, compiler diagnostics, and formal verification results~\cite{wang2024large,wang2024enhancing,liu2024openllm}. Building on this progress, future work could investigate applying RL techniques to the VeriCoder dataset, using the accompanying test cases as a feedback signal to iteratively improve RTL code generation quality.
\section{Conclusion}

Recent advances in Large Language Models (LLMs) have opened new possibilities for Electronic Design Automation (EDA), particularly in RTL code generation. However, most existing datasets emphasize syntactic validity while overlooking functional correctness, which limits the effectiveness of fine-tuned models. We introduce \textsc{VeriCoder}, a model fine-tuned on a dataset with 125,000 examples that is validated for functional correctness. This dataset is constructed using a feedback-directed refinement pipeline guided by a teacher LLM, which generates and iteratively updates both RTL designs and unit tests until the design passes simulation. The resulting dataset consists of functionally validated triples comprising a natural language specification, an RTL implementation, and a passing test. Fine-tuned on this dataset, \textsc{VeriCoder} achieves state-of-the-art results on two established RTL benchmarks, yielding relative improvements of up to 71.7\% on VerilogEval and 27.4\% on RTLLM. An ablation study confirms the impact of functional validation on model performance, underscoring the importance of high-quality training data. Future work may explore formal verification and reinforcement learning to further advance AI-assisted hardware design.

\section*{Acknowledgment}
We thank Samantha Archer, Yao Hsiao, Mohammad Rahmani Fadiheh and Subhasish Mitra for their discussions. This work was partially supported by a Google Research Award.

\bibliographystyle{IEEEtran}
\bibliography{references}

\end{document}